\documentclass[10pt]{iopart}
\usepackage{color}
\usepackage{graphicx}
\usepackage{hyperref}
\usepackage{import}
\usepackage{siunitx}
\usepackage{subfig}

\begin{document}

\rapid[Open-Source Measurement System]{An Open-Source Data Acquisition System for Laboratory and Industrial Scale Applications}
\author{Konstantin Niehaus$^1$ and Andreas Westhoff$^1$}
\address{$^1$German Aerospace Center, Bunsenstraße 10, D-37073 Göttingen, Germany}
\ead{\mailto{konstantin.niehaus@dlr.de}}

\begin{abstract}
	We introduce a cost-efficient, and open-source measurement system for measuring and monitoring indoor air quality,
    aerosol concentration, dust contamination, and comfort-relevant quantities. 
    The system allows to access these quantities simultaneously at multiple sites. 
    The software architecture is described along with the hardware set-up including an overview of the supported measurement transducers so far. 
    The system allows flexible adaptation on laboratory-scales as well as on large-scales like aircraft cabins, vehicles, buildings, and passenger compartments in general.
\end{abstract}
\submitto{\MST}

\ioptwocol

\section{Introduction}
There is a plethora of measurement systems for the examination of air properties, e.g. humidity, aerosol-concentration, thermal comfort, or air quality.
However, many of such systems support just one measurement quantity, are closed-source, cost intensive, or not extensible \cite{Martin2014, Grinias2016, ClarosMarfil2016}.
In addition, the use of costly measurement equipment is an impediment for measurements on large-scales, where a large number of sensors is needed.
This is especially true not only for the research community, but also for a wider public.
Citizen and open science projects aim to reduce barriers, to open findings to a wider audience \cite{OGrady2016} and to allow the reproduction of experiments for a reasonable balance of costs and benefits.
Examples of ongoing projects are monitoring drinking water quality \cite{Brouwer2018}, light pollution \cite{Kyba2013}, or air quality \cite{English2020}.
These topics are of vital interest since they influence our everyday life.
This also applies to indoor air-quality, especially against the background of the SARS-CoV-2 pandemic.
To investigate ventilation concepts inside buildings, the interior of vehicles, and aircraft cabins under realistic environmental conditions, a portable measurement system is needed.
To meet the requirements of measurement application in the afore listed environments, we developed an open and cost-effective multi-sensor measurement platform.
We utilize this system in a wide range of experimental facilities, which vary in their physical dimensions, operational conditions, number of sensors and sensor types.
For instance, we use the system: 

- To assess the risk of infection during the rise of the pathogen SARS-CoV-2 in 2020.
This was achieved by studying the aerosol propagation in aircraft cabins \cite{NLR2021}.
The system's flexible design and low power consumption enabled online in-flight measurements on notebooks driven solely by power banks in airplane passenger compartments of an \textit{Airbus A320}, a \textit{Boeing B737} and a \textit{Boeing B787}.
The use of over 70 optical particle-matter sensors enabled the estimation of infection risk derived from the aerosol concentration and dynamics of artifical saliva in the vicinity of passenger faces.
Investigations of aerosol distributions in a generic classroom-configuration \cite{Lange2022} and the simulation of human respiration in trains \cite{Kohl2021} were also conducted using this system.

- To quantify passenger comfort in compartments resembling aircrafts\ \cite{Lange2020} and trains\ \cite{Schmeling2021} with the objective to investigate novel ventilation concepts.
Therefore, we use our measurement system to quantify energy efficiency, air quality, and passenger comfort.
These multidimensional quantities require the spatial and temporal measurement of i.e. temperature, humidity and CO$_2$ concentration. 

- To investigate the latent and sensible heat transfer in mixed convection with phase transition it is of utmost importance to measure the water vapor content of humid air precisely \cite{Niehaus2021}. 
Here, the system's \textsc{API} aids the control of temperatures, flow rates and humidity and also records measurement values.
Additionally, we use the measurement system's support of high accuracy dew-point mirrors to automatically calibrate capacitive humidity-sensors.

In the following we introduce the system's hardware and software architecture as well as the software interface.

\section{System architecture}
Several open-source and open-hardware based measurement systems have been proposed in the past.
Most of these systems require manufacturing of printed circuit boards (\textsc{PCB})~\cite{Martin2014} and a pre-calibration~\cite{Grinias2016}.
Oftentimes, systems are split into multiple hardware systems, where one or more are responsible for measurement data acquisition and others for the human-machine interface (\textsc{HMI}) ~\cite{ClarosMarfil2016}.
Software on more than two different systems increases the initial installation and system maintenance effort.
In the following, we present a well-proven system architecture that requires only a single host device that communicates with digital transducers via standard hardware and software interfaces.
The main objective is to provide a software and hardware solution with minimal effort to procure, install, and operate.

\subsection{Software requirements}
The source code and hardware description of the presented systems are available in a public repository~\cite{Niehaus2022}.

Both, our in-house code and third-party software packages are open-source and suited for commercial and scientific applications.
A sensor device supported by this system must use an open communication protocol and have a public command-set to evaluate its readings and operational status.
Peripherals and wiring are standard, readily available parts or feature an open design of the \textsc{PCB} used.
The broad accessibility of spare parts or the ability to manufacture them yourself contributes to reduced downtime.
In addition, a browser-based data visualization and control interface makes the system easy to use.
The scope of application ranges from small laboratory-scale experimental set-ups to large set-ups with a variety of sensors.
This scalability also allows the same evaluation code implemented in the proof-of-concept phase to be used and applied at industrial scale.
The support of a well-defined application programming-interface (\textsc{API}) ensures the integration of the present system into higher-level measurement environments or into specialized visualization frameworks.
Furthermore, an open \textsc{API} offers the option to control one or more copies of this system remotely.
This guarantees easy integration into production chains, where the measured values need to be automatically converted into control units.

\subsection{Hardware set-up} 
\begin{figure*}
	\graphicspath{{./images/hardware/}}
	\def\svgwidth{\textwidth}
	\input{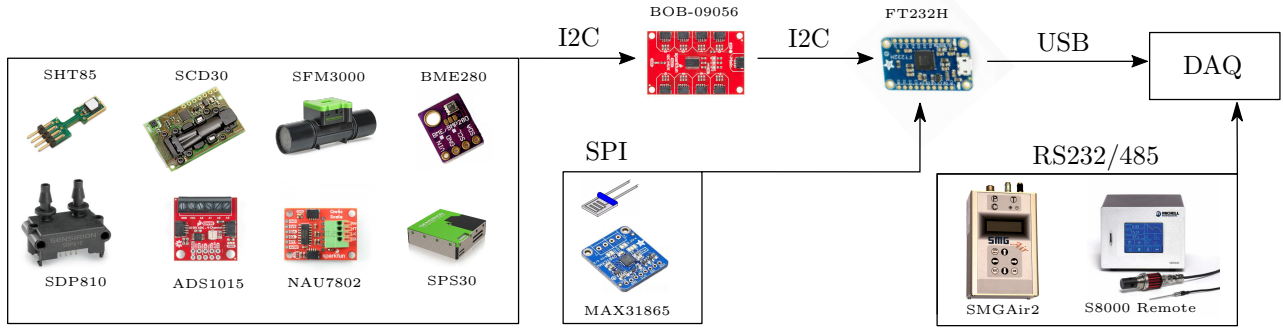}
    \caption{Interface and measurement transducer overview: A \textsc{DAQ} device receives measurement results via the communication interfaces \textsc{SPI}, I$^2$C, \textsc{RS232} and \textsc{RS485}. 
            Before being passed to the \textsc{DAQ} the signals of the \textsc{SPI} and the I$^2$C are connected to \textsc{USB} via a \textsc{FT232H} integrated circuit.
            A multiplexer enables the system to host more than one transducer with the same I$^2$C bus address.
            \textsc{SPI} sensors can be connected to the digital outputs of the \textsc{FT232H}.
    }
	\label{fig:hw}
\end{figure*}
A sketch of the hardware set-up is depicted in \autoref{fig:hw}.
Here, a \textit{Unix}-based single-board computer, a notebook, or a complete desktop computer work as a host system for the data-acquisition system (\textsc{DAQ}, top right).
Device communication is realized via USB and proven standard converters to connect to \textsc{RS232} and \textsc{RS485}.
These two serial bus standards are common in industrial grade measurement devices.
Most low cost measurement transducers rely on Inter Integrated Circuit (I$^2$C) and Serial Peripheral Interface (\textsc{SPI}).
The system provides these interfaces via an \textsc{FT232H} breakout board by \textit{Future Technology Devices International Limited} (\textsc{FTDI}).
Both interfaces require a different wiring concept to support a large number of sensors.
I$^2$C is based on a bus architecture and multiplexing is required whenever there are sensors that respond to identical bus addresses.
This results in a standard data acquisition unit consisting of an eight-channel multiplexer (\textsc{TCA9548A}) and a \textsc{USB}-to-I$^2$C converter.
If there is no electrical current limit, up to 64 sensors of the same type are supported via a single I$^2$C measurement unit.
Due to the bandwidth limit of I$^2$C, the polling rate decreases with increasing sensor counts.
In contrast to I$^2$C, \textsc{SPI} requires one digital output pin per sensor.
Hence, by using a digital multiplexer, up to 16 \textsc{SPI} sensors can be connected to a single USB port.
This is tested using a 4-wire Pt100 temperature measurement box with a digital multiplexer, an \textsc{FT232H}, and 16 \textsc{MAX31865} (4-wire set-up) by \textit{Maxim Integrated} on a single \textsc{PCB} (see \cite{Niehaus2022} for layout).
The system is able to detect whether a measurement unit is configured for \textsc{SPI} or I$^2$C communication.
Plug and play operation is supported by an automated sensor detection. 
The measurement quantities, average measurement accuracies and maximum sampling frequencies of the supported measurement devices are listed in \autoref{tab:sensors}.
It lists only devices and measurement ranges used in referred applications.
However, the list of supported devices is extensive and growing rapidly.
Proven probes and sensor systems will be updated in the project documentation \cite{Niehaus2022}. 

\begin{table}
	\caption{%
        Measurement transducers including accessible measurement quantities (Qty.), 
        maximum measurement frequencies $f_{\max}$, average accuracies, and measurement ranges.
        Quantities are temperature ($T$), relative humidity~($\varphi$), volume-flow rate~($\dot V$), absolute pressure~($P$), differential pressure~($p$), $CO_2$ concentration in air~($c_{co_2}$), dew-point temperature~$T_{dp}$, and voltage~$U$.
        $v$~represents the current measurement value. 
        A full list can be found in the repository~\cite{Niehaus2022}.
    }
	\label{tab:sensors}
	\begin{indented}
		\item[]\begin{tabular}{@{}lllll}
			\br
            Sensor		& $f_{\mathrm{max}}$& Qty.          & Accuracy                      & Range                         \\
			\mr
            SHT85		& \SI{0.5}{\hertz}	& $T$           & $\pm\SI{0.2}{\kelvin}$        & 0   - \SI{80}{\celsius}       \\
                        & 					& $\varphi$     & $\pm\SI{1.5}{\percent}$       & 0   - \SI{80}{\percent}       \\
            ADS1015     & \SI{128}{\hertz}	& $V$           & $\pm\SI{0.25}{\percent}$      & $\pm$\SI{6.1}{\volt}          \\
            MAX31865	& \SI{1}{\hertz}	& $T$           & $\pm\SI{0.1}{\kelvin}$        & -200 - \SI{500}{\celsius}     \\
            SFM3000		& \SI{15}{\hertz}	& $\dot V$      & $\pm0.015v$                   & $\pm$\SI{200}{slm}            \\
            BME280		& \SI{1}{\hertz}	& $T$           & $\pm\SI{0.5}{\kelvin}$        & 0   - \SI{65}{\celsius}       \\
                        & \SI{1}{\hertz}	& $P$           & $\pm\SI{0.1}{\kilo\pascal}$   & 30  - \SI{110}{\kilo\pascal}  \\
                        & \SI{1}{\hertz}	& $\varphi$     & $\pm\SI{3}{\percent}$         & 20  - \SI{80}{\percent}       \\
			SDP810-128  & \SI{0.5}{\hertz}	& $p$           & $\pm\SI{3}{\percent}$         & $\pm$ \SI{125}{\pascal}       \\
                        & 					& $T$           & $\pm\SI{2}{\kelvin}$          & -40 - \SI{85}{\celsius}       \\
            SCD30		& \SI{0.5}{\hertz}	& $c_{co_2}$    & $\pm\SI{30}{ppm}$             & 0.4  - \SI{10}{\kilo ppm}     \\
                        &                   &               & $\ +0.03v$                    &                               \\
                        & 					& $T$           & $\pm\SI{0.6}{\kelvin}$        & 0 - \SI{50}{\celsius}         \\
                        &					& $\varphi$     & $\pm\SI{3}{\percent}$         & 0 - \SI{95}{\percent}         \\
            S8000       & \SI{0.1}{\hertz}	& $T_{dp}$      & $\pm\SI{0.1}{\kelvin}$        & -40 - \SI{125}{\celsius}      \\
                        & 					& $T$           & $\pm\SI{0.1}{\kelvin}$        & -40 - \SI{90}{\celsius}       \\
            SMGAir2     & \SI{30}{\hertz}	& $T$           & $\pm\SI{1}{\kelvin}$          & 4 - \SI{80}{\celsius}         \\
                        & 					& $P$           & $\pm0.02v$                    & -0.8  - \SI{10}{\bar}         \\
                        & 					& $\dot V$      & $\pm0.02v$                    & depending \\
                        & 					&               &                               & on nozzle \\
			\br
		\end{tabular}
	\end{indented}
\end{table}

\subsection{Software and API}
\begin{figure}
	\graphicspath{{./images/software/}}
	\def\svgwidth{0.5\textwidth}
	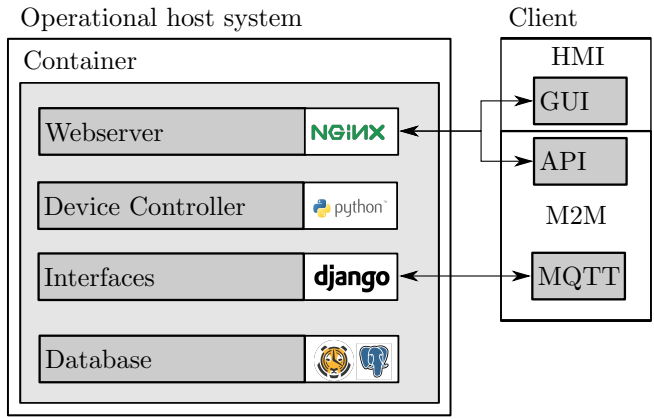
    \caption{%
        Software architecture:
        hosting operational system with a container framework,
        \textit{nginx} web-server, 
        measurement-device controller,
        \textit{django} based interface framework,
        and a \textit{PostgresSQL} time-series database.
        Client server communication: 
        \textsc{HMI} via a webinterface and \textsc{M2M} communication via web \textsc{API} or \textsc{MQTT}.
        }
    \label{fig:sw}
\end{figure}
The web-interface and the sensor drivers are installed inside sandboxed docker containers.
An overview of its structure is listed in \autoref{fig:sw} with the containers, implemented in this project, highlighted in gray.
The instances of the web server, the relational time-series database \textit{TimescaleDB}, data-acquisition workers and the graphical user interface are separated for better maintainability. 
The containers communicate via standardized interfaces, which facilitates the maintenance in case a single container has to be replaced or modified.
Besides configuration adjustments, the database and the \textit{nginx} web-server are imported from their respective repository.
Operational states are represented in two ways and are accessible via \textsc{HTTP}.

\begin{figure}
    \begin{center}
        \subfloat[][Data monitoring]{\includegraphics[width=0.5\textwidth]{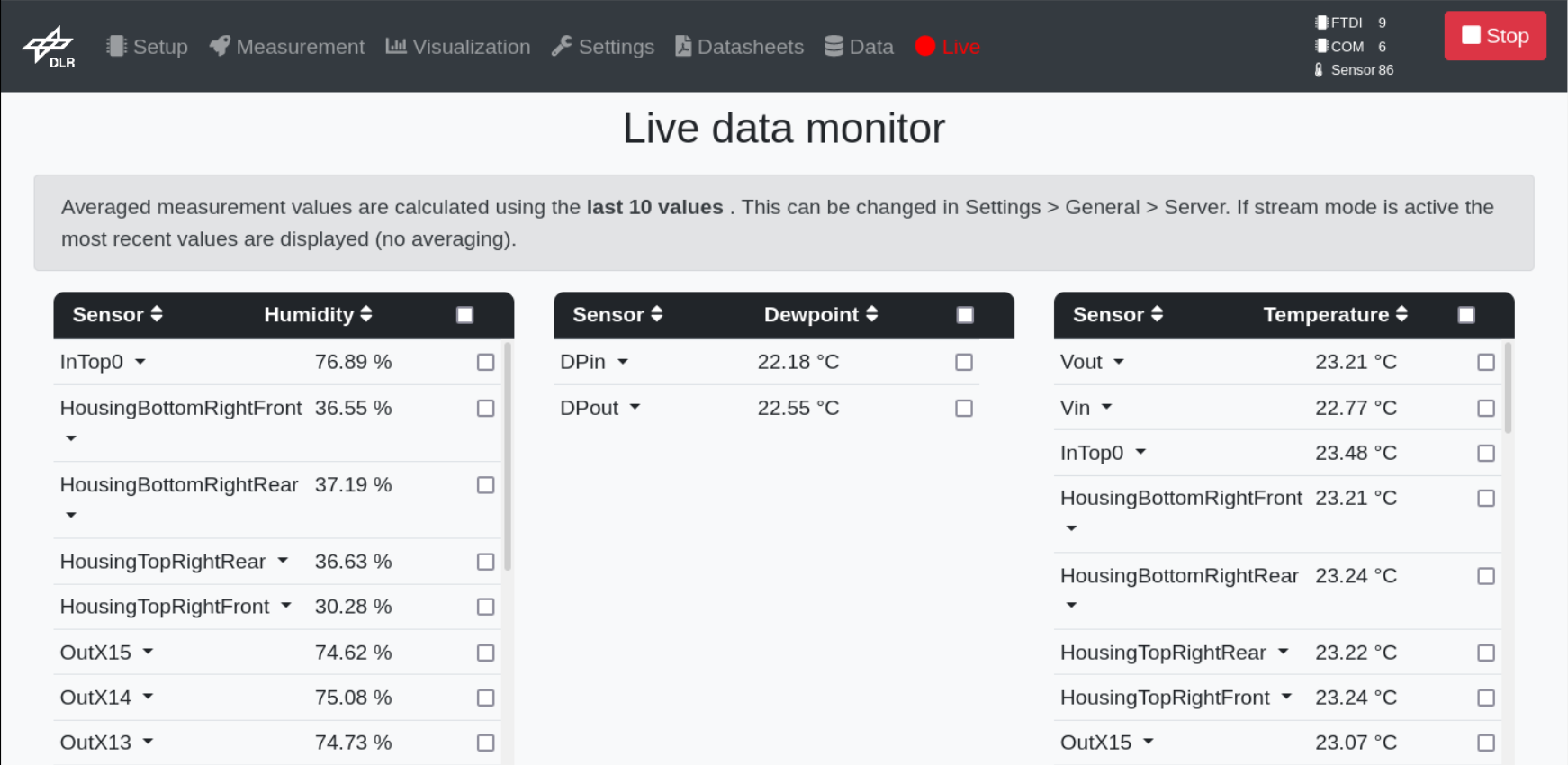}\label{fig:interface_live}}\\
        \subfloat[][Raw data visualization]{\includegraphics[width=0.5\textwidth]{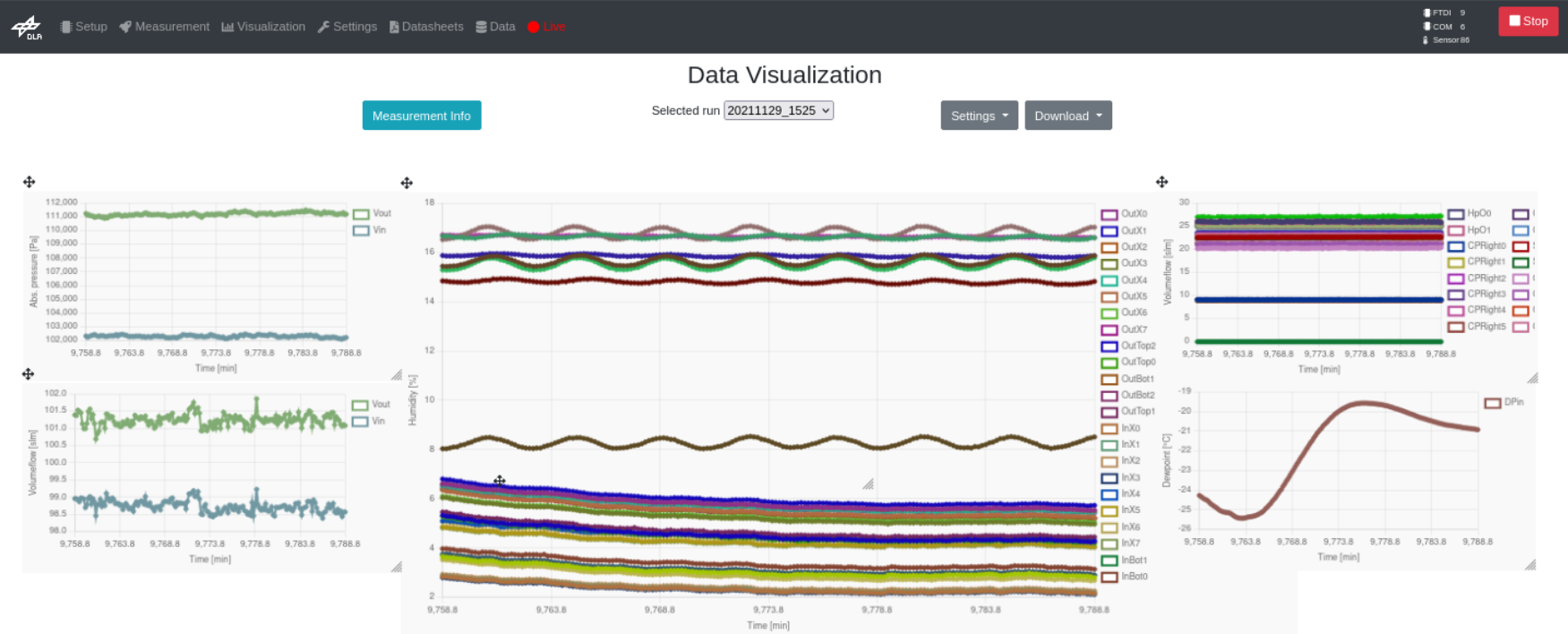}\label{fig:interface_plot}}
    \end{center}
    \caption{Web based \textsc{GUI} for monitorinig and controlling.}
\end{figure}

The first provides a graphical user interface (\textsc{GUI}) to start and stop measurements, to connect new devices, to display real-time data, to visualize results, and to download measurement values.
The second provides an \textsc{API}, which includes and extends the information and control capabilities of the \textsc{GUI} in a machine readable fashion.
The communication with a \textsc{REST}ful \textsc{API} is supported by all state-of-the-art programming languages.
This ensures seamless integration to complement an existing set-up.
In addition, a Message Queuing Telemetry Transport (\textsc{MQTT}) client is used to enable monitoring of data via applications common to the Internet of Things (\textsc{IOT}), such as \textit{Grafana} or \textit{Node-RED}.
\autoref{fig:interface_live} displays the submenu for live data monitoring. 
Further, real-time computation of statistical quantities i.e. the average measurement value of a sensor set is supported.
\autoref{fig:interface_plot} reveals the submenu of the customizable visualization of the probe time-series.
The GUI supports an automatic sensor initialization. 
In addition, the export to ascii and binary based file formats is supported via the web interface or the \textsc{API}. \\

The system stores known sensors identified via unique identification numbers in a database. 
This numbers link to additional information like: calibration data, displaying settings, or sensor positions.
Each sensor has its own optimized initialization routine to ensure stable operational conditions.
The initialization runs before each start of a measurement.
Each measurement unit runs in its own thread to obtain measurement values in parallel.
This also increases the system stability, since connection losses or device failures only affect a single device and have no impact on the system's integrity. 
The system has been tested against disruptive factors like increased vibration, high temperature gradient, or strong electromagnetic interference, where sensors often become unresponsive, or temporarily shut down.
In these cases the system attempts to reinitialize missing sensors and resumes measurements after successful reconnection.
Critical parts of the software have been written using test-driven development to ensure that readings are retrieved, converted and stored correctly.

\section{Conclusion and future development}
We introduced a versatile and compact measurement system.
Features of the system are a modular and easily extensible hardware and software architecture and an open-source and open-hardware approach.
Support of high precision sensors together with low priced measurement transducers allows for applications ranging from low budget citizen science projects to complex systems i.e. precision control loops in fundamental research.
The system has proven itself stable and flexible in a variety of applications.
Future work is dedicated to allow for wireless sensor data transmission and to support \textsc{TCP}/\textsc{IP} measurement devices. 
We cordially invite readers to participate in the development and testing of the system.


\begin{thebibliography}{10}

\bibitem{Martin2014}
F.J.~Ferrero Mart{\'{\i}}n, M.~Valledor Llopis, J.C.~Campo Rodr{\'{\i}}guez,
  J.R.~Blanco Gonz{\'{a}}lez, and J.~Men{\'{e}}ndez Blanco.
\newblock Low-cost open-source multifunction data acquisition system for
  accurate measurements.
\newblock {\em Measurement}, 55:265--271, September 2014.

\bibitem{Grinias2016}
James~P. Grinias, Jason~T. Whitfield, Erik~D. Guetschow, and Robert~T. Kennedy.
\newblock An inexpensive, open-source {USB} {A}rduino data acquisition device
  for chemical instrumentation.
\newblock {\em Journal of Chemical Education}, 93(7):1316--1319, June 2016.

\bibitem{ClarosMarfil2016}
Luis~J. Claros-Marfil, J.~Francisco Padial, and Benito Lauret.
\newblock A new and inexpensive open source data acquisition and controller for
  solar research: Application to a water-flow glazing.
\newblock {\em Renewable Energy}, 92:450--461, July 2016.

\bibitem{OGrady2016}
Michael~J. O'Grady, Conor Muldoon, Dominic Carr, Jie Wan, Barnard Kroon, and
  Gregory M.~P. O'Hare.
\newblock Intelligent sensing for citizen science.
\newblock {\em Mobile Networks and Applications}, 21(2):375--385, January 2016.

\bibitem{Brouwer2018}
Stijn Brouwer, Paul van~der Wielen, Merijn Schriks, Maarten Claassen, and Jos
  Frijns.
\newblock Public participation in science: The future and value of citizen
  science in the drinking water research.
\newblock {\em Water}, 10(3):284, March 2018.

\bibitem{Kyba2013}
Christopher C.~M. Kyba, Janna~M. Wagner, Helga~U. Kuechly, Constance~E. Walker,
  Christopher~D. Elvidge, Fabio Falchi, Thomas Ruhtz, Jürgen Fischer, and
  Franz Hölker.
\newblock Citizen science provides valuable data for monitoring global night
  sky luminance.
\newblock {\em Scientific Reports}, 3(1), May 2013.

\bibitem{English2020}
Paul English, Heather Amato, Esther Bejarano, Graeme Carvlin, Humberto Lugo,
  Michael Jerrett, Galatea King, Daniel Madrigal, Dan Meltzer, Amanda
  Northcross, Luis Olmedo, Edmund Seto, Christian Torres, Alexa Wilkie, and
  Michelle Wong.
\newblock {P}erformance of a {L}ow-{C}ost {S}ensor {C}ommunity {A}ir
  {M}onitoring {N}etwork in {I}mperial {C}ounty.
\newblock {\em Sensors}, 20(11):3031, May 2020.

\bibitem{NLR2021}
{{Netherlands Aerospace Center}}.
\newblock Corsica final report: Quantitative microbial risk assessment for
  aerosol transmission of sars-cov-2 in aircraft cabins based on measurement
  and simulations.
\newblock Technical report, Netherlands Aerospace Centre NLR, June 2021.

\bibitem{Lange2022}
Pascal Lange, Andreas Westhoff, Andreas Kohl, and Axel M{\"u}ller.
\newblock Experimental study of the indoor aerosol-dynamics for a low-momentum
  ventilation system with an air purifier unit.
\newblock {\em SSRN}, pages 1--43, February 2022.
\newblock This manuscript was submitted to Journal of Aerosol Science and is
  currently under review. This is a pre-print version at SSRN.

\bibitem{Kohl2021}
Andreas Kohl, Pascal Lange, and Daniel Schmeling.
\newblock Experimental simulation of the human respiration.
\newblock In {\em Notes on Numerical Fluid Mechanics and Multidisciplinary
  Design}, pages 472--482. Springer International Publishing, July 2021.

\bibitem{Lange2020}
Pascal Lange, Daniel Schmeling, Tobias Dehne, Axel Dannhauer, Felix Werner, and
  Ingo Gores.
\newblock New long-range cabin mock-up enabling the simulation of flight cases
  my means of tempered fuselage elements.
\newblock In {\em Aerospace Europe Conference - AEC2020}, February 2020.

\bibitem{Schmeling2021}
Daniel Schmeling, Tim Berlitz, Thorsten Tielkes, Tobias Dehne, and Marcel
  J{\"a}ckle.
\newblock Demonstrator f{\"u}r {I}nnovationen im {R}eisendenkomfort und
  {K}limatisierung - {DIRK} (in {G}erman).
\newblock In {\em 18. Internationale Schienenfahrzeugtagung Dresden}, pages
  1--3, September 2021.

\bibitem{Niehaus2021}
Konstantin Niehaus, Andreas Westhoff, and Claus Wagner.
\newblock Characterization of a mixed convection cell designed for phase
  transition studies in moist air.
\newblock In {\em New Results in Numerical and Experimental Fluid Mechanics
  XIII}, pages 483--493. Springer International Publishing, July 2021.

\bibitem{Niehaus2022}
Konstantin Niehaus.
\newblock Mobile measurement system https://doi.org/10.5281/zenodo.6471388,
  April 2022.

\end{thebibliography}


\section*{References}
\end{document}